\documentclass{emulateapj}
\usepackage{apjfonts}
\usepackage{psfig}
\def\gs{\mathrel{\raise0.35ex\hbox{$\scriptstyle >$}\kern-0.6em 
\lower0.40ex\hbox{{$\scriptstyle \sim$}}}}
\def\ls{\mathrel{\raise0.35ex\hbox{$\scriptstyle <$}\kern-0.6em 
\lower0.40ex\hbox{{$\scriptstyle \sim$}}}}

\def\etal{\hbox{et al.}$\,$}
\def\Hd{\hbox{H$\delta$}$\,\,$}
\def\Ha{\hbox{H$\alpha$}$\,$}
\def\Hb{\hbox{H$\beta$}$\,$}

\def\Msun{\rm{\hbox{M$_{\odot}$}}}           

\def\kms{\rm{\hbox{km s$^{-1}$}}}
\def\OII{\hbox{[O II]}$\,$}

\def\Hd{\hbox{H$\delta$}$\,$}
\def\24m{\hbox{24~$\micron$}$\,$}

\begin{document}

\slugcomment{Submitted: 2009 April 2; accepted: 2009, May 29}

\shorttitle{Dressler et al.}
\righthead{SFR \& Mode since $z = 0.7$}

\title{Evolution of the Rate and Mode of Star Formation in Galaxies since $z = 0.7$}

\author{Alan Dressler\altaffilmark{1}, Augustus Oemler, Jr.\altaffilmark{1}, 
Michael G. Gladders\altaffilmark{2}, Lei Bai\altaffilmark{1}, Jane R.\ Rigby\altaffilmark{1}, \& Bianca M.\ Poggianti\altaffilmark{3}}

\altaffiltext{1}{The Observatories of the Carnegie Institution of Washington, 813 Santa Barbara St., Pasadena, California 91101-1292}
\altaffiltext{2}{Department of Astronomy \& Astrophysics, University of Chicago, Chicago, IL 60637} 
\altaffiltext{3}{INAF-Osservatorio Astronomico di Padova, vicolo 
dell'Osservatorio 5, 35122 Padova, Italy}

\begin{abstract}

We present the star formation rate (SFR) and starburst fraction (SBF) for a sample of field galaxies from the ICBS 
intermediate-redshift cluster survey.  We use \OII\ and Spitzer \24m fluxes to measure SFRs, and \24m fluxes 
and \Hd \,absorption to measure of SBFs, for both our sample and a present-epoch field sample from the
Sloan Digital Sky Survey (SDSS) and Spitzer Wide-area Infrared Extragalactic (SWIRE) survey.  We find a 
precipitous decline in the SFR since $z = 1$, in agreement with other studies, as well as a corresponding rapid decline 
in the fraction of galaxies undergoing long-duration moderate-amplitude starbursts.  We suggest that the change in 
both the rate and mode of star formation could result from the strong decrease since $z = 1$ of gas available for 
star formation.

\end{abstract}


\section{Introduction}

In recent years, much attention has been focused on the early history of star formation, in particular, 
the rise of the SFR to its peak around $z \sim 2$  (e.g., Fall, Charlot, \& Pei 1996, Madau \etal\,1996, 
Giavalisco \etal\,2004, P{\'e}rez-Gonz{\'a}lez \etal\,2008, Bouwens \etal\,2008).  Such a rapid rise is not surprising 
considering the dynamical timescales of galaxy-sized structures and the lifetimes of stars.   More surprising, 
perhaps, is the precipitous decline in the global SFR since $z \approx 1$ --- more than a factor of 10 lower than 
its peak value, and falling fast (e.g., Lilly \etal\ 1996, Schiminovich \etal\,2005, Hopkins \& Beacom\,2006, 
Villar \etal\,2008). This is remarkable not only for its rapidity but also because it appears to mark our epoch as 
the beginning of the end of galactic star formation. It is now clear that the unexpected prevalence of starforming 
galaxies in rich galaxy clusters discovered by Butcher \& Oemler (1978), long regarded as a cluster phenomenon, 
is universal, as starforming galaxies --- great and small --- surrender youthful vigor and fade towards 
oblivion in only a few billion years.

Using the wide field of the Inamori-Magellan Areal Camera and Spectrograph (IMACS) on Magellan-Baade, the 
{\it IMACS Cluster Building Survey} (ICBS) is focused on the study of galaxy infall and evolution from
$R\sim5$ Mpc into cluster cores.   Because the projected density of cluster/supercluster  members is low at such large 
radii, our near-complete samples necessarily include $\sim$1000 ``field" galaxies at redshift $0.2 < z < 0.8$ per survey 
field.  This gives us an opportunity to  compare galaxy evolution in clusters with the field over this 
epoch.  From these data we report in this {\it Letter} on the significant decline in SFR and SBF since $z\sim1$.
 
\section{Data} 

The data discussed here come from 4 fields that contain rich galaxy clusters at $z = 0.33, 0.38,  0.42,$ $\&\, 0.55$.  
The IMACS f/2 spectra have an observed-frame resolution of 10\,\AA\ full-width-half-max with a typical S/N $\sim$ 20-30 
in the continuum per resolution element. Spectral coverage varies, but almost all cover \OII\ and \Hb\  emission, and \Hd\  
absorption, our optical diagnostics of star formation;  a fraction cover \Ha\ as well.  In each 28\arcmin-diameter IMACS field 
we have observed 65\% of the galaxies that are brighter than $r \sim 22.5$, obtaining adequate spectra of 81\% of these. 
Measurement of spectral features followed procedures described in Dressler \etal (2004, hereinafter D04). For two fields 
we have confusion-limited Spitzer Multiband Imaging Photometer (MIPS) \24m images (Guest Observer program 
40387) that cover nearly the entire IMACS fields. Details of the data, data reduction, and analysis of the field sample are 
described in Oemler \etal\,(2009b, hereinafter Oem09b). 

In the following analysis we include all galaxies between $0.10 < z < 0.70$ with absolute AB magnitudes at 4400\,\AA\  
brighter than $M_{44}^* +1.0$, except for those with redshifts within $\pm 3 \sigma$ of each targeted cluster. $M_{44}^*$ 
has been determined to evolve with redshift as $M_{44}^* = -20.00-z$ (Oem09b); the limit of $M_{44}^* +1.0$ is about equivalent to 
$r = 22.5$ at $z = 0.70$.  There are 1144 objects in this sample. Galaxies are given weights proportional to the inverse of 
the incompleteness of the data at each apparent magnitude. To provide a low--redshift point we use two samples of 
SDSS galaxies with redshifts $0.04 < z < 0.08$. To compare to infrared-derived properties of the ICBS sample, we use a 
sample of 385 SDSS galaxies within the SWIRE fields (Lonsdale \etal\,2003). For comparison with the optically-derived 
properties of our sample, we use this SDSS/SWIRE sample and add 690 SDSS galaxies near the North Galactic Pole. 
We assume a concordance cosmology with $\Omega_{matter} = 0.27$ and $H_o = 71\, \kms\ Mpc^{-1}$.

\section{Measurements of SFR since $z=0.7$}

For galaxies with Spitzer-MIPS coverage we calculate SFRs in a way similar to P{\'e}rez-Gonz{\'a}lez \etal\,(2006), using the \24m flux to 
estimate the absorbed UV flux and \OII\ to estimate the escaped UV flux, but including the k--corrections of Rieke \etal \,(2008) and a 
SFR scale based on a Salpeter initial-mass-function (IMF, vid.\,Oem09b).  Donley \etal \,(2008) report that 10--15\% of \24m sources at 
these flux densities are AGN-dominated.  Given the strong redshift evolution of the AGN luminosity function, the contamination in our 
$z<0.7$ sample should be should be considerably lower.  Figure 1 (bottom panel) shows the median SFR per unit $L/L^*$, which we 
shall refer to as the specific SFR, as a function of redshift. We choose the {\it median}  specific SFR over the more commonly reported 
{\it mean}, because it is provides a more stable and typical value --- particularly at higher redshifts, where the mean is dominated by a small 
number of very luminous objects.  (In the $0.60 < z < 0.70$ interval, only 1.6\% of the objects contribute 56\% of the total star formation.) 
We also show, for comparison, values of the median $SFR/(L/L^*)$, derived from the mid-IR data of Damen \etal\,(2009), and from the 
P{\'e}rez-Gonz{\'a}lez \etal\,(2008) study of the buildup of stellar mass (from near-IR measurements), as reinterpreted by Damen \etal\ and 
converted by us to a Salpeter IMF scale.  To convert the Damen \etal\ values of SFR per unit mass to $SFR/(L/L^*)$ we take a mean 
value of $M/L$ at 4400\,\AA\  of 3, which is the ratio of the cosmic mass density (P{\'e}rez-Gonz{\'a}lez, \etal \,2008) to the cosmic 
luminosity density (Blanton \etal\,2003).  Although cosmic means, both are dominated by galaxies in the luminosity 
and mass ranges of our sample, and thus appropriate for our use.  With $M/L = 3$, a sample with our absolute magnitude limit 
has a mean mass of $5x10^{10}\,\Msun$.

\vspace*{0.6cm}


\vspace*{0.1cm}
\hbox{~}

\centerline{\psfig{file=fig1.eps,angle=0.,width=2.2in}}
\noindent{\scriptsize
\addtolength{\baselineskip}{-3pt}
 
\hspace*{0.3cm}Fig.~1.\ (Bottom) {\it Median} Specific SFR, in solar masses per year per L* luminosity.  Black filled circles 
show values for our sample derived from Spitzer-MIPS \24m fluxes.  Red stars are data from Damen \etal (2009),  
and red open circles are data from P{\'e}rez-Gonz{\'a}lez \etal \,(2008) as reinterpreted by Damen \etal (2009). The median SFR appears
to be declining more rapidly towards the present epoch.  (Top) Median values (solid circles) and 90th percentile values (open circles) 
of \OII for the $z > 0.2$ ICBS field sample, binned to achieve comparable of numbers of \OII-detected galaxies.   The single point 
at $z=0.06$ comes SDSS galaxies with optical spectroscopy (see text). 

\vspace*{0.2cm}
\addtolength{\baselineskip}{3pt}
}

At $z>0.4$ the {\it median} values of SFR increase more slowly than the mean SFR, as reported by Damen \etal 
(2009), Martin \etal (2007), Zheng \etal (2007), but are otherwise qualitatively similar. However, what has not been 
apparent from previous observations is that the fall of the median SFR appears to be accelerating in recent epochs.  

Optical emission lines \OII, \Hb, and \Ha\ are less reliable SFR indicators than \24m flux because of the very 
large and variable extinction which blankets the HII regions of starforming galaxies. However, they do 
provide at least a qualitative measure of current star formation and, unlike our Spitzer-MIPS data, are available 
for all four ICBS fields. Therefore, we plot in the top panel of Figure 1 the 50th and 90th percentiles of the distribution 
of {\it equivalent width} EW(\OII) as a function of redshift.  The shapes of these distributions are qualitatively similar to
the infrared--derived SFR, however, \OII strength increases more slowly with redshift because of the well-known 
effect that galaxies with higher SFRs also have higher dust extinction.

\section{Were Starbursts the Normal Mode of Star Formation before $z=1$?}

It is now well established that starburst and post--starburst galaxies are abundant in intermediate--redshift clusters 
(Poggianti \etal 1999 (hereinafter P99), Oemler \etal 2009a), however, less is known about the prevalence of starbursts 
among field galaxies at these epochs.  Starburst indicators include exceptionally strong Balmer absorption, intense
optical emission, and excess \24m flux.   Strong Balmer absorption in the integrated light of galaxies signals a rapid 
decline in the SFR, because light from A stars persists after the blue continuum from O and B stars, and the emission 
from HII regions --- both of which dilute the Balmer absorption lines --- begin to fade or disappear altogether (P99).
Although strong Balmer absorption accompanies even the simple truncation of star formation, as O and B stars
and their HII regions evolve away, stellar $EW(\Hd) \gs 5$\AA\ occurs only in the aftermath of a significant rise in the SFR 
above its past average, i.e., a burst (Dressler \& Gunn 1983, Couch \& Sharples 1987, P99). Using this criterion, 
P99, D04 and Oemler \etal (2009a) found that the SBF in the field at $z \sim 0.4$ is lower than in clusters, 
but significantly higher than it is in the field today; however, the data sets used were quite small. More recently, Poggianti 
\etal \,(2009) find a large incidence of dusty starburst galaxies for $0.4 < z < 0.8$ in all environments, but particularly in groups.  
Using an entirely different approach, Bell \etal\ (2005) find a substantial fraction of massive field galaxies at $z \sim 0.7$ with SFRs 
much higher than their long--term averages. On the other hand, Noeske \etal \,(2007) argue that the narrow width of the SFR/mass 
versus mass relation among field galaxies in the Groth strip of DEEP2 data is inconsistent with a large fraction of strong starbursts.

Determining the starburst frequency --- SBF --- as a function of cosmic epoch and environment is important both for understanding the 
nature and cause of starbursts, and their role in galaxy evolution. For example, starbursts provide a quick means of exhausting 
the gas supply in a galaxy.  The dense environment of rich clusters provides a variety of mechanisms for producing starbursts, 
including ram-pressure from the intergalactic medium, tidal encounters between galaxies, and mergers and accretion.  Some of 
these are also viable in the group environment, to more or less effect, but ram pressure, for example, is likely to be unimportant.  Some 
present-epoch, truly isolated galaxies appear to have experienced starbursts as well, which might point to some instability in the 
disks of star forming galaxies, or accretion of a small satellite, as other possible causes.
  
Figure 2 compares results from two different methods of determining the SBF. The top panel shows the fraction of
all galaxies in our field sample that have EW(\Hd\,) of 2.5\AA\ or greater than expected for a normal, non--bursting galaxy, 
following the method of D04, but using an improved EW(\Hd) versus EW(\OII) relation (q.v. Oem09).  Following Bell \etal \,(2005), 
we also calculate the SBF from the Spitzer--derived SFRs, by defining a starburst galaxy as one whose observed 
SFR is significantly higher than its long-term average rate, which should be, ignoring mass loss during stellar evolution,

\begin{equation}
SFR_{past average} = M_{gal} / t_{gal} = (L_{gal} \times M/L) / t_{gal}
\end{equation}

\noindent where $t_{gal}$ is the length of time that a galaxy has been forming stars, and $M_{gal}$, $L_{gal}$, and $M/L$ are, 
the galaxy's {\em stellar mass}, luminosity, and stellar mass--to--light ratio. We define $t_{gal}$ to be time elasped 
from redshift $z=4$ (a reasonable assumption for the start of star formation) to the observed redshift, and assume that 
$M_{gal}/L_{gal} = 3$, where $L_{gal}$ is calculated at 4400\,\AA. We identify a galaxy as a starburst if 
$SFR_{obs}/SFR_{past-average} > 3$.  We plot in Figure 2 the fraction of galaxies which meet this criterion and perform the 
same analysis for the observations of Noeske \etal \,(2007) and Bell \etal\,(2005). For these we include objects with 
$M_{gal} \ge 1.6 \times 10^{10}M_{\sun}$, which for $M/L=3$, is equivalent to our luminosity limit of $M_{44}^* +1.0$.

\vspace*{-0.3cm}


\vspace*{1.0cm}
\hbox{~}
\centerline{\psfig{file=fig2.eps,angle=0.,width=2.2in}}
\noindent{\scriptsize
\addtolength{\baselineskip}{-3pt}
 
\hspace*{0.3cm} Fig.~2.\ Two measures of starburst fraction, SBF, versus redshift for galaxies brighter than M*+1. 
Top: fraction of galaxies with excess EW(\Hd) $>$ 2.5\AA;  bottom: fraction of galaxies with SFR $> 3\times 
L_{tot} \times (M/L)$/$\tau$(universe). Black filled circles --- ICBS data; open red circles- data from Noeske \etal (2007); 
red filled circles --- data from Bell \etal (2005).  Black open circles are ICBS data with SFR $> 10\times L_{tot} \times 
(M/L)$/$\tau$(universe).  Using this criterion for a starburst, all three samples give consistent results.

\vspace*{0.2cm}
\addtolength{\baselineskip}{3pt}
}

\noindent

Figure 2 shows that such starbursts make up $\sim$20-25\% of starforming galaxies by $z\sim0.6$, a much larger 
fraction than today,  Moreover, because the ``duty cycle" (fraction of the time these galaxies are identifiable as starbursts) 
is almost certainly less than 50\%, the clear implication is that {\it most starforming galaxies at $z > 0.6$ have undergone 
a starburst of at least moderate strength.}  Figure 2 also shows the SBF for a factor-of-ten increase rising in parallel. 

Noeske \etal \,(2007) emphasize the rarity of starbursts over a comparable (but somewhat wider) redshift range, basing  
their argument on the narrowness of the SFR/M versus M relation.  They conclude that no more than one-third of typical
starforming galaxies have SFR variations greater than a factor-of -two, while our sample suggests, after correction for the
fraction of passive galaxies, that one-third have experienced a rise of a factor of three or more.  Put in these terms, the two 
results appear mildly inconsistent, however, as  Figure 2 illustrates, the Noeske \etal\ data, when analyzed using the same 
method as we and Bell \etal (2005) have used, is in good agreement with our own.  

Finally, we show another way of assessing the SBF(z), using the composite spectra approach developed in D04.  Here 
we add, weighted by luminosity, all ICBS spectra in each of 5 redshift slices $0.20 < z < 0.70$, and for a present-epoch sample 
of SWIRE galaxies.  We plot in Figure 3 EW(\OII, \Hd) measured from these composite spectra, along with expected values for a 
population of (mostly) continuously-starforming galaxies at the present epoch.  Figure 3 includes expected values calculated 
by D04 from the 2dF \& CNOC2 field surveys (see D04) and also new determinations using the SWIRE data, an 
improvement  because we now have individual \OII distributions (on which these predictions rest) from our 
ICBS sample, for each redshift interval.  The Dressler-Shectman low-redshift cluster and field samples follow the
"continuous star formation" prediction, but the clusters at $z \sim 0.4$ studied in the Morphs collaboration depart, indicating
a higher fraction of starbursts (see D04).

\vspace*{-0.3cm}


\vspace*{0.1cm}
\hbox{~}
\centerline{\psfig{file=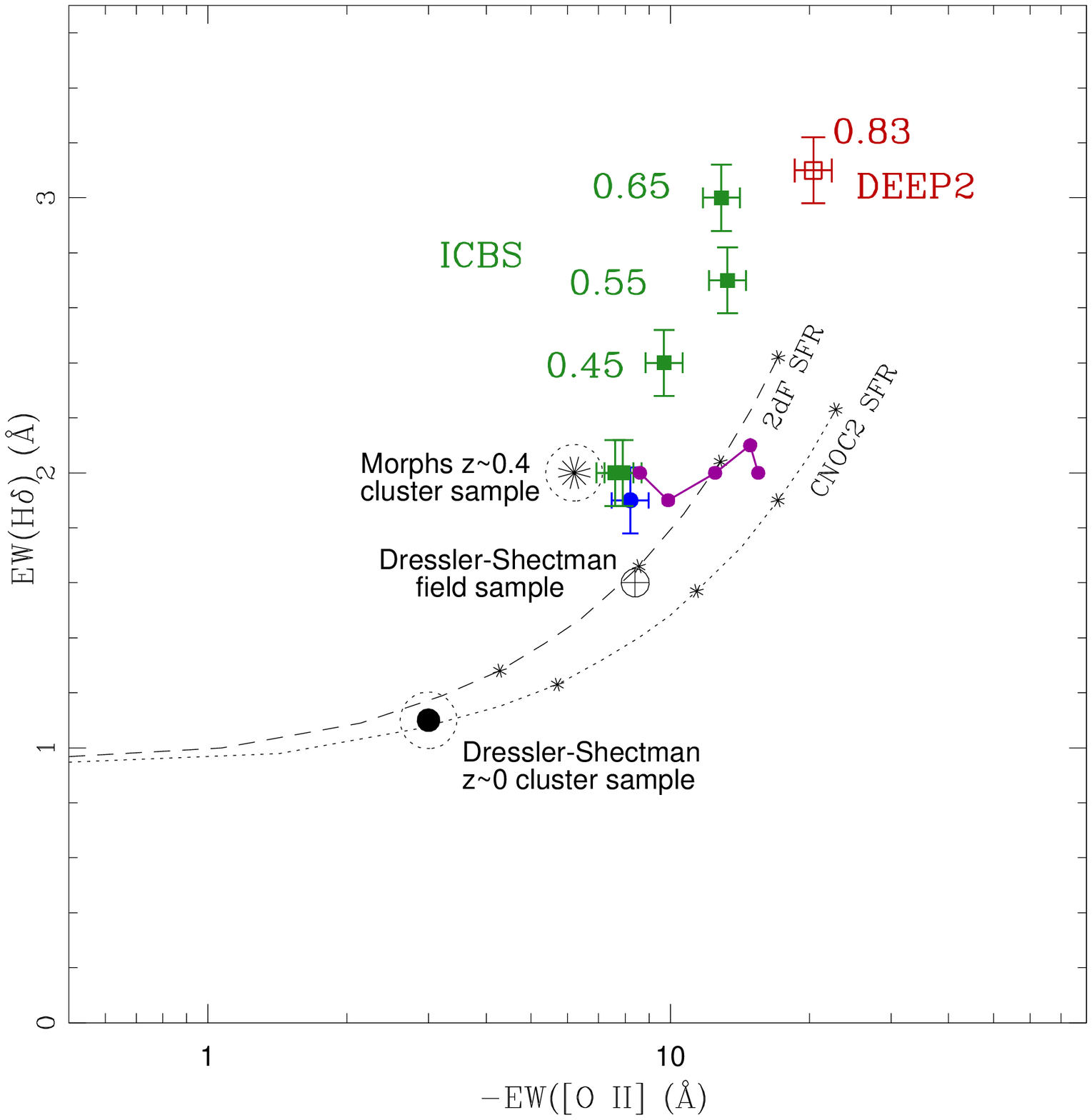,angle=0.,width=3.2in}}
\noindent{\scriptsize
\addtolength{\baselineskip}{-3pt}
\hspace*{0.3cm}Fig.~3.\ \Hd versus \OII equivalent widths for composite spectra, adapted from D04.  The lines `2dF' and 
`CNOC2' show predictions for continuously starforming populations, and the points `Morphs' and `Dressler-Shectman' cluster 
and field samples, from studies described in D04.  The low-redshift SWIRE field galaxy sample is the blue point.  The 
ICBS field sample measurements are the 5 green points for redshift intervals centered at z = 0.25, 0.35, 0.45, 0.55, 0.65.   The 5 joined 
purple points show the prediction of EW(\OII, \Hd) at these 5 redshift intervals for a population of present-epoch starforming 
galaxies (the SWIRE sample) with very few starbursts, as described in the text.   The observed ICBS measurements at $z = 0.45, 55,
 \& 0.65$ far exceed these predicted values, showing the substantial fraction of starburst galaxies in these field populations. The 
DEEP2 (red) point for field galaxies at $z \sim 0.83$ extends and confirms this result.

\vspace*{0.2cm}
\addtolength{\baselineskip}{3pt}
}

For the new field sample, Figure 3 shows that the $z = 0.25, 0.35$ slices are little different from 
the present-epoch population, but the $z = 0.45, 0.55, 0.65$ slices have much stronger \Hd \,absorption than 
expected for a population dominated by continuous star formation.  We also can compare our EW(\OII, \Hd) values to  
those from a composite spectrum of $\sim$1000 field galaxies at $z \sim 0.8$ from the DEEP2 Galaxy Redshift Survey 
(Davis \etal\,2009),  This coadded spectrum was produced as described in Weiner \etal \,(2008) by Jeff Newmann and 
Renbin Yan from galaxies selected with limiting $M_R$ comparable to our own sample.  The DEEP2 point continues the 
trend of yet greater SBF to even higher redshift, for a data set independent of our own, with higher S/N and higher spectral 
resolution. 

In summary, methods based on (1) individual \Hd\ strengths, (2) the SFR-increase-over-past-average, and (3) (EW(\OII, \Hd) 
from composite spectra, all point to an SBF that increases markedly with redshift.

\section{Discussion and Conclusions}

We have used optical spectra and mid-IR fluxes of field galaxies in our ICBS galaxy cluster survey to measure the rapid decline 
of the specific star formation rate, SFR, since $z \sim 0.7$.  Although qualitatively consistent with other studies, we find that 
--- parameterized as the median specific SFR ---  this decline appears to be steepening toward the present epoch.  This
might be evidence for a kind of ``downsizing:" elliptical/bulge formation dominates at $2 < z < 6$ and smoothly transitions 
to the building of massive galaxy disks, but as this wanes for $z < 1$, there is insufficient mass in the still-starforming 
dwarf galaxy population to prevent what is essentially the end of the era of star formation (Dressler 2004).

We also show that it is not just the SFR, but also its mode --- the starburst fraction, SBF --- that is evolving rapidly since $z = 0.7$.  
We use three different methods, from discrete and composite spectra, to show that the SBF also declines dramatically over 
this period.  When accounting for the ``duty cycle," the data suggest that the {\it majority} of starforming galaxies at $z \gs 0.5$ 
followed this mode of star formation rather than in the steadier, ``continuous" mode that dominates today. We emphasize that, 
unlike short and often-intense nuclear starbursts, these are typically of moderate strength --- the SFR rising by a factor of 3--10.  
They are likely to be galaxy-wide and of longer duration  --- consistent with the dynamical timescales of the larger region 
involved.

It seems possible that these two effects --- the rapid decline in specific SFR and the changing mode of 
star formation --- share a common cause.  Availability of gas for star formation could be the principal agent.  
If the Schmidt-Kennicutt (Kennicutt 1998) law holds, higher SFRs of intermediate-redshift galaxies were due to gas 
fractions higher than the 20-40\% typical of today's luminous spirals (McCaugh \& de Blok 1997), although how much 
greater may depend on the relative importance of $H_{2 }$ versus $HI$ (Robertson \& Kravtsov 2009, Krumholz \etal \,2009).  
As discussed by Putnam \etal (2008), this greater gas content could be due to initial supply, accretion of gas-rich satellites, 
or resupply from surrounding reservoirs of cooling gas, the latter the subject of much recent discussion (see, e.g., 
Keres \etal\ 2009).  However, direct measurements of gas contents for intermediate-redshift  galaxies are not yet 
possible, and absorption measurements by Prochaska and Wolfe (2009) --- albeit of HI alone --- support a different 
picture in which gas fractions do not evolve strongly and high SFRs are instead supported by high rates of gas accretion.

If higher SFRs were the result of higher gas fractions of more than 50\% for typical $z > 0.5$ galaxies, then it is reasonable to 
suppose that the higher SBF might also be a result.  Star formation in gas-rich disks may be unstable if high supernovae rates 
heat disks sufficiently to disrupt conditions favorable for star formation.  After a gas-cooling time of several hundred million years, 
rapid star formation could return; color-magnitude diagrams for some nearby dwarf galaxies show two or more episodes of star 
formation separated by billions of years (e.g, Gallart \etal \,1999; Held \etal \,2000).  If resupply by cold gas flows is important to 
the evolution of spiral disks, the higher infall rates of earlier epochs could be subject to significant variability, simply a result of the 
granularity of the gas density in a complex network of filaments.  

Finally, mergers of gas-rich galaxies might contribute to the rising SBF with increasing redshift, although Bell \etal \,(2005) find only 
a small fraction of their high SFR galaxies at $z\sim0.7$ in major mergers. Perhaps a higher rate of accretion of small satellites 
at earlier times might have resulted in more starbursts.  Indeed, since these smaller systems were probably gas rich, accretion rather 
than major mergers might be the dominant starburst trigger.

As yet there are no direct means to measure gas fractions in $z\sim 1$ galaxies, but ALMA and the proposed SKA 
will in future provide that capability.  Until that time, there is an important role for numerical modeling, with its increasing 
resolution, to explore whether gas fractions, inflow rates, and satellite accretion would affect not just the rate of star 
formation but its mode as well.

\section{Acknowledgments}

Dressler and Oemler acknowledge the support of the NSF grant AST-0407343.  All the authors
thank NASA for its support through NASA-JPL 1310394.  Jane Rigby is supported by a Spitzer Space
Telescope Postdoctoral Fellowship.  Partial support was also provided through contract 1255094
from JPL/Caltech to the University of Arizona.  The authors gratefully acknowledge the Jeff Newman, 
Renbin Yan, and the DEEP2 team for the composite spectrum used in this paper.

{\it Facilities:} \facility{Spitzer (MIPS)}, \facility{Magellan:Baade (IMACS)}, \facility{Magellan:Clay (LDSS3)}

\end{document}